\documentclass[aps,prb,reprint,superscriptaddress]{revtex4-1}
\usepackage{graphicx}
\usepackage{amsmath,amssymb}
\usepackage[T1]{fontenc}%.......... Type 1 outline fonts
\usepackage{textcomp}%............. Additional text symbols
\usepackage{txfonts}%.............. Roman-Times, Sans-Helvetica, TT-Monospaced
\usepackage{bm}%................... Bold math fonts
\usepackage{epstopdf}
\usepackage[colorlinks]{hyperref}

\usepackage{xcolor,soul}
\usepackage{ulem}

\begin{document}

\title{Tunable dimensional crossover and magnetocrystalline anisotropy in Fe$_2$P-based alloys}

\author{I. A. Zhuravlev}
\affiliation{Department of Physics and Astronomy and Nebraska Center for Materials and Nanoscience,
University of Nebraska-Lincoln, Lincoln, Nebraska 68588, USA}
\author{V. P. Antropov}
\affiliation{Ames Laboratory, Ames, Iowa, 50011, USA}
\author{A. Vishina}
\affiliation{Kings College London, London WC2R 2LS, UK}
\author{M. van Schilfgaarde}
\affiliation{Kings College London, London WC2R 2LS, UK}
\author{K. D. Belashchenko}
\affiliation{Department of Physics and Astronomy and Nebraska Center for Materials and Nanoscience,
University of Nebraska-Lincoln, Lincoln, Nebraska 68588, USA}

\date{\today}

\begin{abstract}
Electronic structure calculations are used to examine the magnetic properties of Fe$_2$P-based alloys and the mechanisms through which the Curie temperature and magnetocrystalline anisotropy can be optimized for specific applications. It is found that at elevated temperatures the magnetic interaction in pure Fe$_2$P develops a pronounced two-dimensional character due to the suppression of the magnetization in one of the sublattices, but the interlayer coupling is very sensitive to band filling and structural distortions. This feature suggests a natural explanation of the observed sharp enhancement of the Curie temperature by alloying with multiple elements, such as Co, Ni, Si, and B. The magnetocrystalline anisotropy is also tunable by electron doping, reaching a maximum near the electron count of pure Fe$_2$P. These findings enable the optimization of the alloy content, suggesting co-alloying of Fe$_2$P with Co (or Ni) and Si as a strategy for maximizing the magnetocrystalline anisotropy at and above room temperature.
\end{abstract}

\maketitle

Transition-metal pnictide alloys based on Fe$_2$P have attracted considerable attention due to the unusual sensitivity of their magnetic properties to temperature, pressure, external magnetic field, and alloying, \cite{rev91,Fruchart,Jain,exp1,exp2} as well as their possible magnetocaloric applications,\cite{rev91} and they have been extensively studied theoretically. \cite{Costa2012,DelczegCzirjak_2,DelczegCzirjak,DelczegCzirjak2010} Of particular interest is the magnetocrystalline anisotropy (MCA) of Fe$_2$P, which, at 2.3 MJ/m$^3$, is record-high at zero temperature for systems without heavy elements. \cite{Tujii1974,Fujii1978} Although ferromagnetism in Fe$_2$P vanishes through a first-order phase transition at $T_C=216$ K, this temperature can be greatly increased by alloying with Si, Ni, Co, and other easily available elements. \cite{rev91,Fruchart,Chandra1980,Jernberg1984,Catalano1973} The origin of such unusual $T_C$ sensitivity to alloying is not understood. In combination with large MCA and appreciable coercivity observed \cite{Vos1962} in Fe$_2$P alloyed with Co, this feature makes Fe$_2$P-based alloys interesting for permanent-magnet applications.

Here we use first-principles calculations to study the magnetic properties of Fe$_2$P-based alloys. We propose that $T_C$ is easily tunable thanks to the two-dimensional (2D) character of the exchange interaction developing at elevated temperature. We also find that MCA is controlled largely by band filling and is maximized close to the electron count corresponding to pure Fe$_2$P. Based on these findings, we argue that co-alloying with Co (or Ni) and Si is the optimal strategy to maximize the MCA at and above room temperature.

We use the Green function-based formulation of the tight-binding linear muffin-tin orbital method in the atomic sphere approximation. \cite{Turek} Substitutional disorder was treated using our implementation of the coherent potential approximation (CPA) \cite{CPA} with spin-orbit coupling (SOC) included and the MCA energy $K$ calculated following Refs.\ \onlinecite{Turek2008,Belashchenko2015,Zhuravlev2015}. The atomic sphere radii were carefully chosen to reproduce the full-potential band structure. Exchange and correlation were treated within the generalized gradient approximation.\cite{PBE} A uniform mesh of $12\times12\times20$ points provided sufficient accuracy for the Brillouin zone integration.

At $x<0.2$ the (Fe$_{1-x}$Co$_x$)$_2$P alloy has a hexagonal structure with lattice constants that are almost independent of $x$. \cite{Fruchart} Therefore, we used the experimental values for pure Fe$_2$P ($a=5.8675$ \AA, $c=3.4581$ \AA)\cite{Fruchart} at all concentrations in these alloys, and the internal coordinates from Ref.\ \onlinecite{Carlsson1973}.

There are two inequivalent Fe sites in Fe$_2$P: the tetrahedral, weakly magnetic Fe$_\mathrm{I}$ and the pyramidal, strongly magnetic Fe$_\mathrm{II}$.\cite{exp1, DelczegCzirjak} Co and Ni have a strong tendency to occupy the Fe$_\mathrm{I}$ site.\cite{Fruchart,note-site} The results we report here assume 100\% site preference, but there is no qualitative difference with equal substitution on both types of sites.

Using the linear-response formalism, \cite{LAG1987} we find, in agreement with earlier results, \cite{DelczegCzirjak_2} that the exchange parameters in the ferromagnetic state do not change much with 10-15\% substitution of Co, Ni, or Si, while the experimental $T_C$ increases sharply. For example, $T_C$ nearly doubles at 10\% Co substitution, while the \textit{paramagnetic} Curie temperature $\theta_P$ increases only by 14\%.\cite{Kumar-Co} The dominant exchange parameters are strong ferromagnetic Fe$_\mathrm{II}$-Fe$_\mathrm{II}$ and a weaker but comparable Fe$_\mathrm{I}$-Fe$_\mathrm{II}$. Mean-field theory predicts $T_C$ of order 700 K that depends weakly on concentration, consistent with the behavior of $\theta_P$. Thus, the dramatic influence of various alloying elements on $T_C$ is quite puzzling. It was suggested \cite{DelczegCzirjak} that the stabilization of the Fe$_\mathrm{I}$ local moments at elevated temperature is responsible for the $T_C$ increase in Si-doped Fe$_2$P, but it is unclear how this mechanism would apply to Co and Ni doping which have an opposite effect on band filling.

We propose the following scenario, which is consistent with first-principles calculations and experimental evidence. We observe that the magnetic structure of Fe$_2$P consists of alternating layers of Fe$_\mathrm{I}$ and Fe$_\mathrm{II}$ sites. The Fe$_\mathrm{I}$ sites are weakly magnetic, and their exchange coupling to Fe$_\mathrm{II}$ sites is considerably weaker than the in-plane Fe$_\mathrm{II}$-Fe$_\mathrm{II}$ coupling, while the  Fe$_\mathrm{I}$-Fe$_\mathrm{I}$ exchange is negligibly small. Therefore, we expect that the Fe$_\mathrm{I}$ sublattice magnetization declines much faster than Fe$_\mathrm{II}$ as the temperature approaches $T_C$. Indeed, a neutron diffraction measurement found a very small Fe$_\mathrm{I}$ magnetization just below the first-order $T_C$ in 7\% Ni-substituted Fe$_2$P.\cite{Kumar-ND} As a result, the Fe$_\mathrm{I}$-mediated coupling between the Fe$_\mathrm{II}$ layers is strongly suppressed near $T_C$. On the other hand, as we will now show, direct interlayer Fe$_\mathrm{II}$-Fe$_\mathrm{II}$ coupling is weak in pure Fe$_2$P and strongly sensitive to band filling. Overall, this behavior indicates a crossover from three-dimensional to quasi-two-dimensional magnetism as a function of both temperature and composition of the alloy.

\begin{figure}[htb]
\includegraphics[width=0.85\columnwidth]{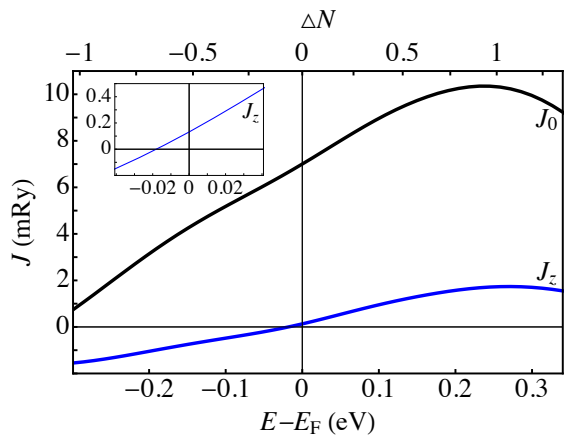}
\caption{Dependence of exchange interaction on Fermi level position in paramagnetic Fe$_2$P alloys between pyramidal Fe sites. Black: total exchange interaction ($J_0$); blue: interlayer exchange between two nearest layers of  pyramidal Fe$_\textrm{II}$ sites ($J_z$). Inset: enlargement of $J_z$ at small doping. Top axis: band filling $\Delta N$.}
\label{fig:J0JzFermi}
\end{figure}

To calculate the direct coupling between the Fe$_\mathrm{II}$ layers, we model the paramagnetic state using the disordered local moment approach and calculate the Fe$_\mathrm{II}$-Fe$_\mathrm{II}$ exchange parameters $J_{ij}$ using the linear response formalism. \cite{Gyorffy1985} The local moments on the Fe$_\mathrm{I}$ site vanish in this approach. We focus on the total effective exchange $J_0=\sum_j J_{ij}$ and the interlayer exchange $J_z=\sum^\prime_j J_{ij}$, where the prime restricts the summation to sites $j$ in the Fe$_\mathrm{II}$ layer that is adjacent to the one containing site $i$.
Fig.\ \ref{fig:J0JzFermi} shows $J_0$ and $J_z$ calculated as a function of the Fermi energy $E_F$ in the rigid-band model; \cite{reviewsd} the upper axis shows the band filling $\Delta N$, which is the electron count per formula unit, referenced from pure Fe$_2$P. It is seen that $J_0$ increases smoothly by about 30\% at $\Delta N\sim1$, which corresponds to a 50\% substitution of all Fe atoms by Co. In a real alloy, $J_0$ is not expected to increase as much, because alloying with Co reduces the magnetic moments. On the other hand, although $J_z$ shows a similar trend, it is small at $\Delta N\approx0$ and changes sign at a small hole doping (see inset in Fig.\ \ref{fig:J0JzFermi}).

This result suggests that the magnetic structure effectively becomes quasi-2D at elevated temperatures. In this scenario, $T_C$ can be strongly suppressed from its mean-field value, \cite{2DHeis} while nothing special happens to $\theta_P$. Moreover, Fig.\ \ref{fig:J0JzFermi} shows that the degree of two-dimensionality is sensitive to electron doping, and we expect the effect of alloying or another perturbation on $T_C$ should correlate with its effect on $J_z$. While the electron (hole) doping makes the exchange interaction less (more) two-dimensional, specific alloying elements can also affect $J_z$ in ways that are unrelated to electron count, such as through the structural distortions induced by the size effect.

\begin{figure}[htb]
\includegraphics[width=0.85\columnwidth]{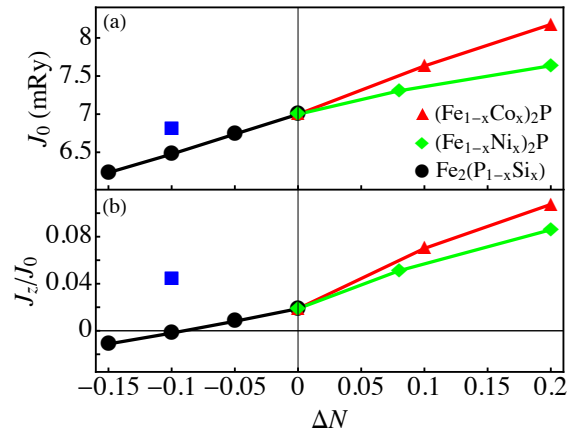}
\caption{Dependence of the exchange coupling on band filling $\Delta N$ in paramagnetic Fe$_2$P-based alloys. (a) Total exchange interaction $J_0$. (b) Ratio of interlayer exchange between two nearest layers of  pyramidal Fe sites and $J_0$. Symbols connected by lines: at experimental lattice parameters for Fe$_2$P. Blue squares: at experimental lattice parameters \cite{Jernberg1984} for Fe$_2$P$_{0.9}$Si$_{0.1}$.}
\label{fig:Fe2PJz}
\end{figure}

Going beyond the rigid-band model, Fig.\ \ref{fig:Fe2PJz} shows $J_0$ and the $J_z$/$J_0$ ratio computed using CPA in paramagnetic Fe$_2$P alloyed with Co, Ni, or Si, taking substitutional \cite{LAG1987} and spin \cite{Gyorffy1985} disorder on the same footing. If the lattice constants are kept fixed (solid lines), the CPA results are similar to the predictions of the rigid-band model. However, it turns out the lattice distortion induced by Si overwhelms its effect on the electron count, and overall Si increases $J_z$ (blue square). On the other hand, the structural distortion induced by alloying with Co or Ni is negligible. Thus, Co, Ni, and Si \textit{all} increase $J_z$, despite their opposite effects on the band filling. All these alloying elements also sharply increase $T_C$, which supports the idea that the degree of magnetic two-dimensionality correlates with $T_C$ and is otherwise hard to explain, given the opposite (and small) effects of Co and Si on $J_0$.

\begin{figure}[htb]
\includegraphics[width=0.85\columnwidth]{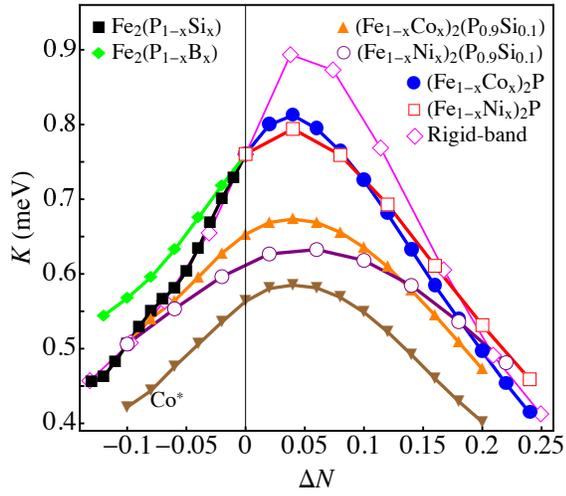}
\caption{Dependence of $K$ on band filling $\Delta N$ in different alloys (see legend) calculated in CPA. Inverted brown triangles (labeled Co$^*$): (Fe$_{1-x}$Co$_x$)$_2$(P$_{0.9}$Si$_{0.1}$) with experimental lattice parameters \cite{Jernberg1984} for Fe$_2$(P$_{0.9}$Si$_{0.1}$). Diamonds: rigid-band approximation for Fe$_2$P.}
\label{fig:dopedSiMaeHex}
\end{figure}

We now turn to magnetocrystalline anisotropy. Fig.\ \ref{fig:dopedSiMaeHex} shows the dependence of $K$ in Fe$_2$P on the electron doping in the rigid-band model, as well as the CPA results for various alloys, all plotted as a function of the electron count $\Delta N$. First, we focus on three-component alloys with Co, Ni, Si, or B. As in the case of the exchange coupling, the rigid-band approximation agrees quite well with CPA calculations for these alloys. This agreement suggests that the MCA energy in Fe$_2$P is not dominated by spin-orbit ``hot spots,'' which would be strongly suppressed by disorder. \cite{Zhuravlev2015} Further, we see that the behavior of $K$ is primarily controlled by band filling. For example, $K$ behaves very similar in (Fe$_{1-x}$Co$_x$)$_2$P and (Fe$_{1-x}$Ni$_x$)$_2$P alloys when plotted against $\Delta N$, which means that the effect of Ni is equivalent to twice as much Co. The effect of Si or B also fits closely with the band filling trend.

The small offset of the maximum in Fig.\ \ref{fig:dopedSiMaeHex} is likely an artifact of the atomic sphere approximation, which slightly overestimates the exchange splitting. While there is no intrinsic reason for the maximum to occur exactly at $\Delta N=0$, we found that a full-potential rigid-band calculation reproduces the trend seen in Fig.\ \ref{fig:dopedSiMaeHex}, but the maximum occurs almost exactly at $\Delta N=0$. The decline of the MCA energy with the increasing concentration of Ni or Co agrees with the experimental data.\cite{Tujii1974,Fujii1978,Kumar-Co}

To identify the mechanisms of MCA and its dependence on band filling, we first consider the site and spin decomposition of the anisotropy of the SOC energy. \cite{SOPT,Belashchenko2015} We have verified that, as expected, this quantity closely follows the concentration dependence of $K$ in these alloys. The dominant term in the spin decomposition comes from the mixing of the minority-spin states of Fe$_\mathrm{II}$ by the $\hat S_z\hat L_z$ term in SOC, while the majority-spin and the spin-off-diagonal term are fairly small in the interesting range of concentrations. The contribution of Fe$_\mathrm{I}$ behaves similarly but is a few times smaller. Given that the diagonal minority-spin contribution dominates, the MCA is approximately proportional to the orbital moment anisotropy,\cite{book1,SOPT} which may be easier to measure.\cite{book1}

\begin{figure}[htb]
\includegraphics[width=0.85\columnwidth]{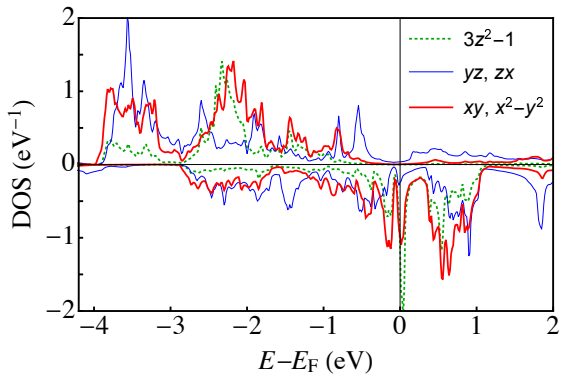}
\caption{Orbital-resolved partial density of $3d$ states for a Fe$_\mathrm{II}$ atom.}
\label{fig:HexPdDOS}
\end{figure}

The orbital-resolved density of states for a Fe$_\mathrm{II}$ atom in pure Fe$_2$P (Fig.\ \ref{fig:HexPdDOS}) shows two closely-spaced peaks of the $xy$ and $x^2-y^2$ character (or $m=\pm2$ in the $Y_{lm}$ basis) separated by the Fermi level. Mixing of these states by the $\hat L_z$ operator leads to the MCA maximum seen in Fig.\ \ref{fig:dopedSiMaeHex}.

\begin{figure}[htb]
\includegraphics[width=0.85\columnwidth]{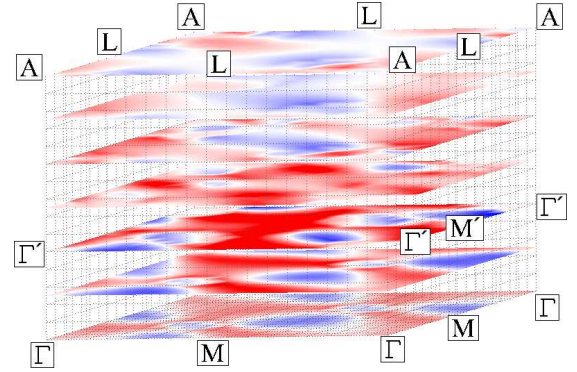}
\caption{Brillouin zone map of the $\mathbf{k}$-resolved minority-spin contribution to $K$ for pure Fe$_2$P. Color intensity indicates the magnitude of positive (red) or negative (blue) contributions. The $\Gamma^{\prime}$ and M$^{\prime}$ points are at $k_z=|\Gamma A|/3$ on the $\Gamma$A and ML lines, respectively.}
\label{fig:kresolvedMaeHex}
\end{figure}

We now examine the contributions to $K$ coming from different regions in the Brillouin zone. \cite{Belashchenko2015}
Fig.\ \ref{fig:kresolvedMaeHex} shows the difference of the minority-spin single-particle energies for magnetization along the $x$ and $z$ axes in pure Fe$_2$P, resolved by $\mathbf{k}$.
We see that the MCA accumulates over a fairly large part of the Brillouin zone, with the most important contribution coming from the region with $k_z\approx k_{1/3}=|\Gamma A|/3$ (bright red area in Fig.\ \ref{fig:kresolvedMaeHex}). Note that an earlier analysis \cite{Costa2012} focused only on high-symmetry directions in the Brillouin zone, leading to an erroneous conclusion that the dominant contributions to $K$ come from band splittings along the KM and $\Gamma$A lines. This observation underlines the need to examine the contributions coming from the entire Brillouin zone.

\begin{figure}[htb]
\includegraphics[width=0.85\columnwidth]{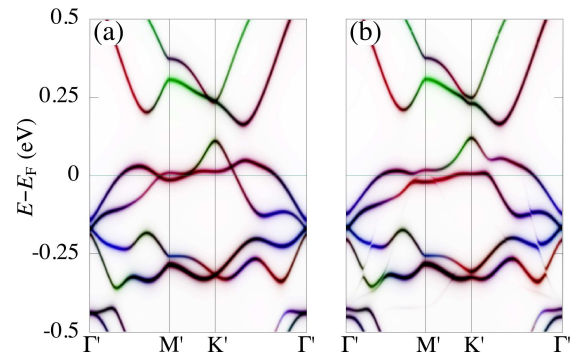}
\caption{Partial minority-spin spectral function in the $k_z=|\Gamma A|/3$ plane for the transition-metal site in Fe$_2$P. (a) Without SOC. (b) With SOC for $\mathbf{M}\parallel z$. The intensities of the red, blue, and green color channels are proportional to the sum of $m=\pm2$ ($xy$ and $x^2-y^2$), sum of $m=\pm1$ ($xz$ and $yz$), and $m=0$ ($z^2$) character, respectively.}
\label{fig:specFunHex}
\end{figure}

The origin of the dominant positive contribution from the vicinity of the $k_z\approx k_{1/3}$ plane can be understood by examining the partial minority-spin spectral function for the transition-metal site. Fig.\ \ref{fig:specFunHex} shows this spectral function for pure Fe$_2$P, resolving $3d$ orbital contributions by color. We see two intersecting bands with the Fermi level cutting through them. The SOC strongly splits these bands for $\mathbf{M} \parallel z$ [panel (b)], while the splitting is much weaker for $\mathbf{M} \parallel x$ (not shown). Although there is no exact spin-orbital selection rule, the states near the Fermi level are predominantly of the $xy$ and $x^2-y^2$ character (see Fig.\ \ref{fig:HexPdDOS}). These orbitals are strongly mixed by the $\hat L_z$ (but not $\hat L_x$) operator, which gives a positive contribution to $K$.

The dependence of $K$ in (Fe$_{0.85}$Co$_{0.15}$)$_2$P on the $c/a$ ratio is similar to the results of Ref.\ \onlinecite{Costa2012} for pure Fe$_2$P: a 5\% increase in $c/a$ results in a 30\% increase in MCA, while the magnetization is slightly decreased. With increasing volume the MCA increases slightly, at a rate of about 2\% per 1\% volume increase, while the magnetization is nearly constant.

At room temperature the anisotropy field and coercivity in (Fe$_{1-x}$Co$_x$)$_2$P are maximized at certain Co concentrations.\cite{Vos1962} This is because there is a tradeoff between low $T_C$ at low concentrations and low MCA, even at $T=0$, at high concentrations. For permanent-magnet applications, it is of interest to maximize the MCA at the operating temperature. First, we considered the prospects of increasing the MCA of the (Fe$_{0.85}$Co$_{0.15}$)$_2$P alloy by substituting a small amount of Fe by an additional alloying element, such as one with a stronger SOC. We found that Mn, Tc, and Re increase $K$ when equally substituted on both Fe sublattices, with Mn having the largest effect and Re the smallest. For example, a 5\% substitution of Re increases $K$ by about 15\%, and that of Mn by as much as 70\%. However, if Mn fully segregates to the pyramidal site (which it strongly prefers \cite{Catalano1973}), the MCA is only enhanced by 20\%. The effect of Mn is then essentially that of the reduced band filling, which could be achieved by simply reducing the amount of Co. Equal substitution of both Fe sites by Ru and Os has almost no effect on $K$, while Rh and Ir reduce it. Thus, double substitution of Fe by Co and another transition-metal element is not promising.

On the other hand, given that $K$ is sensitive to band filling and is maximized close to the electron count of pure Fe$_2$P, the following strategy suggests itself: co-alloy Fe$_2$P with Co or Ni and Si (on Fe and P sublattices, respectively) to maintain the optimal band filling while increasing $T_C$ well above the operating temperature. For example, as shown in Fig.\ \ref{fig:dopedSiMaeHex}, the MCA in (Fe$_{1-x}$X$_x$)$_2$P$_{0.9}$Si$_{0.1}$ alloys, where X is either Co or Ni, has the same dependence on band filling as the three-component alloys. Although the lattice distortion corresponding to Fe$_2$P$_{0.9}$Si$_{0.1}$ reduces the maximal MCA, it is still achieved at the same band filling.

We have examined the Bloch spectral functions in these alloys and found that, in all cases, all the bands seen in Fig. \ref{fig:specFunHex} are easily identifiable and their broadening is fairly small, while the band filling determines the location of the Fermi level. The band broadening does, however, reduce the MCA in (Fe$_{1-x}$X$_x$)$_2$P$_{0.9}$Si$_{0.1}$ compared to (Fe$_{1-x}$X$_x$)$_2$P at the same band filling. Therefore, for alloys with the optimal band filling, co-alloying with Co (or Ni) and Si still comes with a tradeoff between the increasing $T_C$ and decreasing the MCA at $T=0$.

As mentioned above, the shift of the maximum in $K$ from $\Delta N=0$ is likely due to a slightly overestimated exchange splitting. Thus, we predict that MCA is maximized close to the 1:2 doping ratio for Co and Si, or 1:4 for Ni and Si.

CPA calculations show that the ground-state magnetization in Fe$_2$P alloys with Co, Ni, and Si, in the relevant range of concentrations, is approximately a linear function of the band filling: $M\approx M_0+A \Delta N$, where $A\approx -0.85$ $\mu_B$/f.u. In particular, at $T=0$ the ``optimal'' co-doped alloys should have approximately the same magnetization as pure Fe$_2$P.

Consider the series of (Fe$_{1-x}$Co$_x$)$_2$P$_{1-\bar y}$Si$_{\bar y}$ alloys where $\bar y\approx2x$ maximizes $K$ at $T=0$ for the given $x$. It is of practical interest to find $x$ that maximizes $K(T,x)$ in this alloy at the given $T$. To find this maximum, we calculate $K(0,x)$ in CPA and approximate the temperature dependence as follows.

It is often possible to calculate $K(T)$ using the disordered local moment method, \cite{Staunton,Zhuravlev2015} but we do not have a quantitative model to represent the statistical distribution of the magnetic moments and their orientations in the present system with a weakly magnetic sublattice and a dimensional crossover. Therefore, we turn to the experimental data on $K(T)$, which was measured at several concentrations in Ni-doped Fe$_2$P. \cite{Tujii1974,Fujii1978} The $K(T/T_C)$ curves at 0, 10, and 20\% Ni are monotonic and essentially identical when scaled by $K(0)$. Therefore, we assume that $K(T,x)/K(0,x)=f(T/T_C)$ is the same function of $T/T_C$ at any $x$, and we take it from experiment.\cite{Tujii1974,Fujii1978}

The missing piece is the dependence of $T_C$ on $x$. Consider $T_C(x,y)$ in the alloy where $x$ and $y$ are unrestricted concentrations of Co and Si. Since we are considering the series of alloys at optimal band filling, we expect that the increase of $T_C(x,\bar y)$ with $x$ is primarily due to the increase in $J_z$ due to the structural distortion introduced by Si. In Fe$_2$(P$_{1-y}$Si$_{y}$), the hole doping reduces the effect of this structural distortion. Therefore, we can use $T_C(0,\bar y)$ from the experimental data for Fe$_2$(P$_{1-y}$Si$_{y}$) as the lower bound for $T_C(x,\bar y)$ in the lower-bound estimate for $K(T,x)=f(T/T_C)K(0,x)$.

\begin{figure}[hb]
\includegraphics[width=0.85\columnwidth]{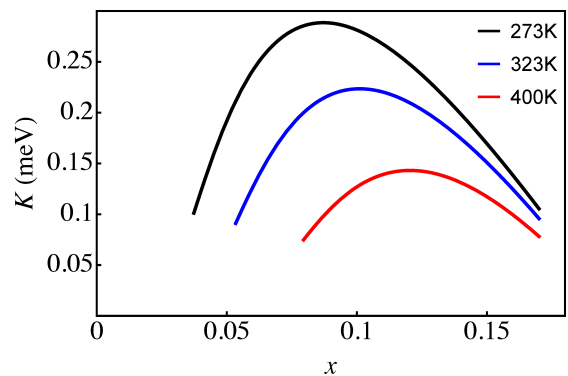}
\caption{Concentration dependence of $K$ at 273K (black), 323K (blue) and 400K (red) in (Fe$_{1-x}$Co$_x$)$_2$P$_{1-\bar y}$Si$_{\bar y}$ alloys with optimal band filling reached at $\bar y\approx2x$. See text for details.}
\label{fig:maeFTemp}
\end{figure}

The resulting lower-bound estimates for $K(T,x)$ are shown in Fig.\ \ref{fig:maeFTemp} for three different temperatures. At room temperature, MCA reaches a maximum of about 0.26 meV/f.u. at $x \approx 0.09$. At higher temperatures this maximum shifts to larger concentrations, while the maximum MCA declines. Thus, correcting for the 2\% shift of the $K(\Delta N)$ curve,
we predict that the optimal alloy for permanent-magnet applications in the 300-400 K operating range has 7-10\% Co (or 3.5-5\% Ni) and a compensating amount of 14-20\% Si. This target composition should be the starting point for experimental verification.
More generally, understanding of the mechanisms by which alloying affects the Curie temperature and MCA provides a path for optimizing the composition of Fe$_2$P-based alloys for specific applications.

The work at UNL was supported by the National Science Foundation through Grants No. DMR-1308751 and DMR-1609776 and performed utilizing the Holland Computing Center of the University of Nebraska. Work at Ames Lab was supported by the Critical Materials Institute, an Energy Innovation Hub funded by the US DOE. Ames Laboratory is operated for the US DOE by Iowa State University under Contract No. DE-AC02-07CH11358. MvS was supported by the EPSRC CCP9 Flagship Project No. EP/M011631/1.


\begin{thebibliography}{99}
\bibitem{note-site} Using VASP, \cite{VASP} we confirmed that Co and Ni prefer the tetrahedral site, by 215 and 75 meV/atom, respectively, compared to the pyramidal site.
\end{thebibliography}
\end{document}